\documentclass[aps,preprint,showpacs,floats,epsf,epsfig,nofootinbib,12pt]{revtex4}
\usepackage{graphicx}
\usepackage{epsfig}
\usepackage[dvips]{color}

\def\beq{\begin{eqnarray}}
\def\eeq{\end{eqnarray}}
\def\non{\nonumber}

\def\la{\langle}
\def\ra{\rangle}
\def\Mb{M_{\mathcal{B}_1}}
\def\Mc{M_{\mathcal{B}_2}}

\begin{document}

\title{ Study on $\Xi_{cc}\to\Xi_c$ and $\Xi_{cc}\to\Xi'_c$  weak decays in the light-front quark model}

\vspace{1cm}

\author{ Hong-Wei Ke$^{1}$   \footnote{khw020056@hotmail.com}, Fang Lu$^{1}$, Xiao-Hai Liu$^{1}$ \footnote{xiaohai.liu@tju.edu.cn} and
        Xue-Qian Li$^2$\footnote{lixq@nankai.edu.cn},
   }

\affiliation{  $^{1}$ School of Science, Tianjin University, Tianjin 300072, China
\\
  $^{2}$ School of Physics, Nankai University, Tianjin 300071, China }

\vspace{12cm}

\begin{abstract}
In this work we study the weak decays of $\Xi_{cc}\to\Xi_c$ and
$\Xi_{cc}\to\Xi'_c$  in the light-front quark model. Generally,
a naive, but reasonable conjecture suggests that
the $cc$ subsystem in $\Xi_{cc}$ ( $us$ pair in $\Xi^{(')}_c$)
stands as a diquark with definite spin and color assignments.
During the concerned processes, the diquark
of the initial state is not a spectator, and must be broken.
A Racah transformation would decompose the original $(cc)q$ into
a combination of $c(cq)$ components. Thus we may deal with the decaying $c$ quark
alone while keeping  the $(cq)$ subsystem
as a spectator. With the re-arrangement of the inner structure
we calculate the form
factors numerically and then obtain the rates of semi-leptonic decays and non-leptonic decays,
which will be measured in the future.

\end{abstract}

\maketitle

\section{Introduction}

In 2017, the LHCb collaboration observed a doubly charmed baryon
$\Xi^{++}_{cc}$\cite{Aaij:2017ueg} with  mass
$3621.40\pm0.72\pm0.27\pm0.14$ MeV which indeed was long-expected
by physicists of high energy physics. In terms of the constituent
quark model there should exist 40 baryon-states of $J^{P}=1/2$ and
35 $J^{P}=3/2$ states which are composed of the five flavors $u$,
$d$, $s$, $c$ and $b$. Most of the light baryons and several heavy
baryons with only one heavy quark ($b$ or $c$) have been observed
experimentally. Thus the doubly heavy baryons would be the goal of
experimental search. One of such states $\Xi^{++}_{cc}$ was
reported by the SELEX\cite{Mattson:2002vu,Ocherashvili:2004hi}
collaboration with a mass about 3520 MeV, however, it was not
confirmed by other collaborations. Recently $\Xi^{++}_{cc}$ has
been measured by the LHCb collaboration in the four-body final
state $\Lambda_cK^{-}\pi^+\pi^+$ \cite{Aaij:2017ueg} and later
$\Xi^{++}_{cc}$ was observed via the $\Xi^{+}\pi^+$
portal\cite{Aaij:2018gfl}.

Theoretically, the structure about the series of  $\Xi_{cc}$ has
not been fully investigated yet (in this paper we only consider
$\Xi^{++}_{cc}$ and the result can be generalized to
$\Xi^{+}_{cc}$ ). For example, its life time and decay rates into
several main channels are not well measured yet while on the
theoretical aspect a reliable approach to study the decays of
doubly heavy baryons is still lacking. Therefore, all attempts to
deeply investigate $\Xi_{cc}$ from different angles should be
valuable. By this study  we may determine the inner structure of
$\Xi_{cc}$ and its decay behaviors by the light front quark model,
consequently the results can be tested and the gained knowledge
would be helpful for designing new experiments for searching other
baryons with two heavy quarks.

Since the mass of $\Xi_{cc}$ is smaller than the production
threshold of $\Lambda_c$ and $D$, it only decays via weak
interaction. Apparently $\Xi_{cc}$ should favorably decays into
products involving a single-charmed baryon. In order to evaluate
its decay rates we first need to know its inner structure. A naive
and reasonable conjecture suggests that the subsystem of the two
$c$ quarks composes a diquark  as a color source for the light
quark \cite{Falk:1993gb,Chang:2007xa}.

In most of works about single charmed-baryons,
the two light quarks can be
regarded as a light diquark\cite{Ebert:2006rp,Korner:1992wi}. Thus while dealing with
weak decays of such baryons, the diquaks with two light quarks can be safely
regarded as spectators which do not undergo any changes during the transitions.
The spectator scenario indeed greatly
alleviates the difficulties of theoretical derivations.

By contraries, in the decay process of $\Xi_{cc}$, one of the two $c$
quarks in initial diquark would transit into a lighter quark by emitting gauge bosons
and the new quark in the final
state will be bound with the spectator $u$ quark to form a light quark subsystem of color anti-triplet
$us$. Namely the picture is that the old diquark  is broken and a new diquark
emerges during the transition.
Anyhow, the diquark (no matter the original ($cc$) or the final
($us$) ) can no longer be treated as a spectator. Therefore the simple
quark-diquark picture could not be reasonably applied to this
decay process.

In this paper we will  extend the
light-front quark model to study the weak decays of $\Xi_{cc}$
where three-body vertex function was obtained in our previous works.
The light-front quark model (LFQM) is a relativistic quark
model which has been applied to study transitions among mesons and
the results agree with the data within reasonable error tolerance
\cite{Jaus,Ji:1992yf,Cheng:1996if,Cheng:2003sm,Hwang:2006cua,Lu:2007sg,
Li:2010bb,Ke:2009ed,Ke:2010,Wei:2009nc,Choi:2007se,
Ke:2009mn,Ke:2011fj,Ke:2011mu,Ke:2011jf}. We also studied the weak
decays of $\Lambda_b$ and $\Sigma_b$ in  the
heavy-quark-light-diquark picture of
baryon\cite{Ke:2007tg,Wei:2009np,Ke:2012wa,Ke:2017eqo} in this
model and our results
\cite{Ke:2007tg,Wei:2009np,Ke:2012wa,Ke:2017eqo} are consistent
with those given in literatures. Thus we have a certain confidence that the
the extension of the light-front quark model to baryon cases is
also successful to the leading order at least
\cite{Wang:2017mqp,Chua:2018lfa,Yu:2017zst,Ke:2007tg,
Wei:2009np,Ke:2012wa,Ke:2017eqo,Ke:2019smy}.

In Ref.\cite{Ke:2019smy} we construct the three-body vertex function
which is applied to  the decay of heavy baryon. Now we try extending the
approach to the concerned process. The transition process can be divided into two steps:
first the old diquark of $cc$ is broken and a subsystem of $cu$ serves as a spectator
during the transition, and then in the finally produced $\Xi_c$, the subsystem of $cu$
would be broken again and a proper structure of $c(us)$ is reformed via the QCD interaction.
As a matter of fact, it is easy
to realize as we  rearrange the ($cc$)diquark-($u$)quark structure into
a combination of  the $cu$(diquark-like subsystem)-$c$(quark) structures by a Racah
transformation. During the transition the ($cu$)diquark-like subsystem can be regarded as
a spectator. Namely, one $c$ quark transits into an $s$
quark but the other $c$ quark and $u$ quark are not touched
approximately. Then for the second step we also need a Racah rearrangement.

In Ref.\cite{Wang:2017mqp,Yu:2017zst} the authors
used the quark-diquark picture to explore the weak decays of doubly
charmed baryon.

Indeed, since in $\Xi_{cc}$ the u-quark has a relative momentum
with respect to the diquark $cc$ (the distant between two $c$
quark is small), thus after the recombination, in the subsystem
$uc$, between the two constituents $u$ and $c$, there exists a
relative momentum. Therefore, rigorously speaking, the subsystem
of $uc$ is a diquark-like subsystem. In our work, we have to take
into account the momenta carried by all the individual quarks
which would undergo some changes during the transition. It is
stressed again that in this work, we treat the combination
involving one $c$ quark and a $u$ quark as a diquark-like
effective subsystem. In other words, $cc$ and $us$ in $\Xi_{cc}$
and $\Xi^{(')}_{c}$ possess definite spin and color quantum
numbers, so we can transform physical subsystem (diquarks) into
effective subsystems. However, since the subsystem $cu$ is not a
diquark, the inner degree of freedom could not be ignored.

 This paper is
organized as follows: after the introduction, in section II we
write up the transition amplitude for $\Xi_{cc}\rightarrow
\Xi^{(')}_{c}$ in the light-front quark model and give the form
factors, then we present our numerical results for
$\Xi_{cc}\rightarrow \Xi^{(')}_{c}$ along with all necessary input
parameters in section III. Section IV is devoted to our conclusion
and discussions.

\section{$\Xi_{cc}\rightarrow \Xi_{c}$ in the light-front quark model}

\subsection{the vertex functions of $\Xi_{cc}\rightarrow \Xi_{c}$}
In our previous works
\cite{Ke:2007tg,Wei:2009np,Ke:2012wa,Ke:2017eqo}, we employ the
quark-diquark picture to study the baryon transitions, where the
diquark has definite spin and serves as a spectator approximately
during the transition process. However in the present concerned
process the picture is no longer valid. For a generally accepted
consideration the two charm quarks in $\Xi_{cc}$ compose a diquark
which stands as a color source for the light $u$ quark which is
moving around with a certain relative momentum with respect to the
diquark. The relative orbital angular momentum between the two $c$
quarks is zero, i.e. the $cc$ pair is in an $S$-wave, due to the
symmetry requirement the spin of the $cc$ pair must be 1. In
Ref.\cite{Ebert:2006rp} the $us$-diquark in $\Xi_{c}$ is  a scalar
diquark whereas in $\Xi'_{c}$ it is a vector. In the decay process
of $\Xi_{cc}$  the $cc$ diquark must be physically broken and one
of the two charm quarks transits into an $s$-quark via weak
interaction and a light $cu$ subsystem is formed which becomes a
spectator for the decay portal. That means neither the diquarks
$cc$ in the initial state  and $us$ in the final state  are
spectators. To realize the transition, we mathematically re-order
the quark structure of $(cc)u$ into a sum of $\sum c(cu)_i$ where
the sum runs over all possible configurations (spin etc. ) via the
Racah transformation. Because of existence of relative momenta
among the quarks, in this work we  explore the baryon transition
in the three-quark picture where the three quarks are individual
subjects and possess their own momenta.

In analog to Refs.\cite{Ke:2019smy,pentaquark1,pentaquark2}  the
vertex functions of $\Xi_{cc}$ and $\Xi_{c}$  with total spin
$S=1/2$ and momentum $P$ are
\begin{eqnarray}\label{eq:lfbaryon}
&& |\Xi_{cc}(P,S,S_z)\rangle=\int\{d^3\tilde p_1\}\{d^3\tilde
p_2\}\{d^3\tilde p_3\} \,
  2(2\pi)^3\delta^3(\tilde{P}-\tilde{p_1}-\tilde{p_2}-\tilde{p_3}) \nonumber\\
 &&\times\sum_{\lambda_1,\lambda_2,\lambda_3}\Psi^{SS_z}(\tilde{p}_i,\lambda_i)
  \mathcal{C}^{\alpha\beta\gamma}\mathcal{F}_{ccu}\left|\right.
 c_{\alpha}(p_1,\lambda_1)c_{\beta}(p_2,\lambda_2)u_{\gamma}(p_3,\lambda_3)\ra,\\
  && |\Xi^{(')}_{c}(P,S,S_z)\rangle=\int\{d^3\tilde p_1\}\{d^3\tilde
p_2\}\{d^3\tilde p_3\} \,
  2(2\pi)^3\delta^3(\tilde{P}-\tilde{p_1}-\tilde{p_2}-\tilde{p_3}) \nonumber\\
 &&\times\sum_{\lambda_1,\lambda_2,\lambda_3}\Psi^{(')SS_z}(\tilde{p}_i,\lambda_i)
  \mathcal{C}^{\alpha\beta\gamma}\mathcal{F}_{csu}\left|\right.
  s_{\alpha}(p_1,\lambda_1)c_{\beta}(p_2,\lambda_2)u_{\gamma}(p_3,\lambda_3)\ra.
\end{eqnarray}

As the spectator approximation cannot be directly applied, dealing
with the process seems more complicated. In fact the $c$ quark
which does not transit via weak interaction and the $u$ quark play
the same role in the transition of $\Xi_{cc}\to \Xi_{c}$, i.e.
they are approximate spectator and their combination can be
regarded as an effective subsystem. Actually the $cc$ and  $us$
are physical subsystems for $\Xi_{cc}$ and $\Xi_{c}$ respectively
since they possess definite spin-color quantum numbers.
By the aforementioned rearrangement of quark flavors the physical
states $(cc)u$  and $c(us)$ are written into sums over effective
forms $c(cu)$ and $s(cu)$ for $\Xi_{cc}$ and $\Xi_c$ respectively.
The detailed transformations are\cite{Wang:2017mqp}
\begin{eqnarray}
&&{[c^1c^2]}_{1}[u]=\frac{\sqrt{2}}{2}(-\frac{\sqrt{3}}{2}[c^2][c^1u]_{0}+\frac{1}{2}[c^2][c^1u]_{1}\nonumber
\\&&\,\,\,\,\,\,\,\,\,\,\,\,\,\,\,\,\,\,\,\,\,\,\,\,\,\,\,\,\,\,\,\,\,\,\,\,\,\,\,\,-\frac{\sqrt{3}}{2}[c^1][c^2u]_{0 }+\frac{1}{2}[c^1][c^2u]_{1})\\
&&[su]_{0}[c]=-\frac{1}{2}[s][cu]_{0}+\frac{\sqrt{3}}{2}[s][cu]_{1}\\
&&[su]_{1}[c]=\frac{\sqrt{3}}{2}[s][cu]_{0}+\frac{1}{2}[s][cu]_{1}
\end{eqnarray}
and then
\begin{eqnarray*}
\Psi^{SS_z}_{ccu}(\tilde{p}_i,\lambda_i)=\sqrt{2}[-\frac{\sqrt{3}}{2}\Psi^{SS_z}_0(\tilde{p}_i,\lambda_i)
+\frac{1}{2}\Psi^{SS_z}_1(\tilde{p}_i,\lambda_i)],\\
\Psi^{SS_z}_{csu}(\tilde{p}_i,\lambda_i)=-\frac{1}{2}\Psi^{SS_z}_0(\tilde{p}_i,\lambda_i)
+\frac{\sqrt{3}}{2}\Psi^{SS_z}_1(\tilde{p}_i,\lambda_i),\\
\Psi^{'SS_z}_{csu}(\tilde{p}_i,\lambda_i)=\frac{\sqrt{3}}{2}\Psi^{SS_z}_0(\tilde{p}_i,\lambda_i)
+\frac{1}{2}\Psi^{SS_z}_1(\tilde{p}_i,\lambda_i),
\end{eqnarray*}
with\cite{Tawfiq:1998nk}
\begin{eqnarray}
\Psi^{SS_z}_0(\tilde{p}_i,\lambda_i)=&&A_0 \bar
u(p_3,\lambda_3)[(\bar
P\!\!\!\!\slash+M_0)\gamma_5]v(p_2,\lambda_2)\bar u(p_1,\lambda_1)
u(\bar P,S) \varphi(x_i,k_{i\perp}),
\end{eqnarray}

\begin{eqnarray}
\Psi^{SS_z}_1(\tilde{p}_i,\lambda_i)=&&A_1 \bar
u(p_3,\lambda_3)[(\bar
P\!\!\!\!\slash+M_0)\gamma_{\perp\alpha}]v(p_2,\lambda_2)\bar
u(p_1,\lambda_1) \gamma_{\perp\alpha}\gamma_{5}u(\bar
P,S)\varphi(x_i,k_{i\perp}),
\end{eqnarray}
where $p_1$ is the the momentum of the $c$-quark which
participates in the transition, $\,p_2\,,p_3$ are the momenta of
the spectator quarks $c$ and $u$, and  $\lambda_1,\lambda_2,
\lambda_3$ are the helicities of the constituents.

Under the normalization of the state $|\Xi_{cc}\rangle$ (or
$|\Xi^{(')}_{c}\rangle$) , \beq\label{A121}
 \la
 \Xi_{cc}(P',S',S'_z)|\Xi_{cc}(P,S,S_z)\ra=2(2\pi)^3P^+
  \delta^3(\tilde{P}'-\tilde{P})\delta_{S'S}\delta_{S'_zS_z}.
 \eeq
and \beq\label{A122}
\int(\prod^3_{i=1}\frac{dx_id^2k_{i\perp}}{2(2\pi)^3})2(2\pi)^3\delta(1-\sum
x_i)\delta^2(\sum
k_{i\perp})\varphi^*(x_i,k_{i\perp})\varphi(x_i,k_{i\perp})=1.
 \eeq

With a simple manipulation, one can obtain\cite{Ke:2019smy}
\begin{eqnarray}
A_0&&=\frac{1}{4\sqrt{P^+(M_0m_1+p_1\cdot\bar{P})(m_2m_3M_0^2+m_3M_0p_2\cdot\bar{P}+
m_2M_0p_3\cdot\bar{P}+p_2\cdot\bar{P}p_3\cdot\bar{P})}}\nonumber\\&&=\frac{1}
{4\sqrt{P^+M_0^3(m_1+e_1)(m_2+e_2)(m_3+e_3)}}.
\\A_1&&=\frac{1}{4\sqrt{3P^+(M_0m_1+p_1\cdot\bar{P})(M_0m_2+p_2\cdot\bar{P})
(M_0m_3+p_3\cdot\bar{P})}}\nonumber\\&&=\frac{1}{4\sqrt{3P^+M_0^3(m_1+e_1)(m_2+e_2)(m_3+e_3)}}.
\end{eqnarray}

The spatial wave function is
 \beq\label{A122}
\varphi(x_i,k_{i\perp})=\frac{e_1e_2e_3}{x_1x_2x_3M_0}\varphi(\overrightarrow{k}_1,\beta_1)
\varphi(\frac{\overrightarrow{k}_2-\overrightarrow{k}_3}{2},\beta_{23})
 \eeq
with
$\varphi(\overrightarrow{k},\beta)=4(\frac{\pi}{\beta^2})^{3/4}{\rm
exp}(\frac{-k_z^2-k^2_\perp}{2\beta^2})$.

\subsection{Calculating the  form factors of $\Xi_{cc}\to\Xi_c$ and $\Xi_{cc}\to\Xi'_c$ in LFQM}

\begin{figure}
\begin{center}
\scalebox{0.8}{\includegraphics{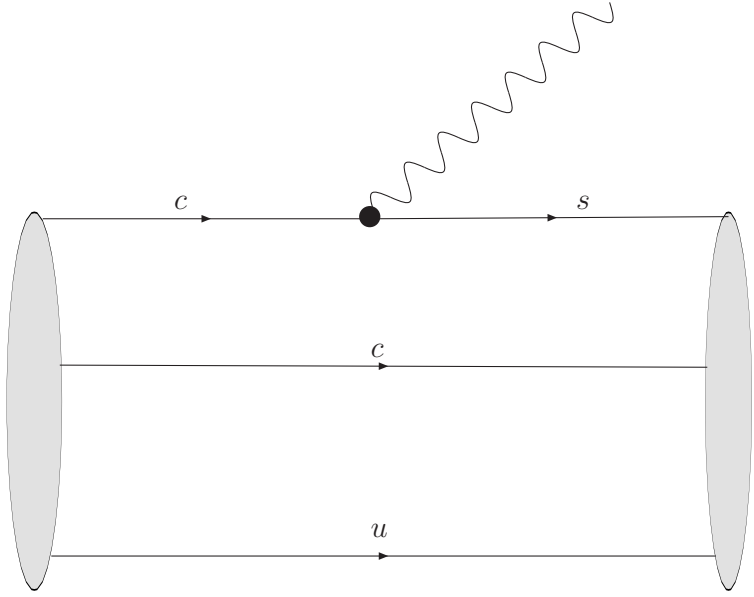}}
\end{center}
\caption{The Feynman diagram for $\Xi_{cc}\to\Xi_{c}^*$
transitions, where $\bullet$ denotes $V-A$ current
vertex.}\label{t1}
\end{figure}

The lowest order Feynman diagram responsible for the weak decay
$\Xi_{cc}\to\Xi_c$ is shown in Fig. \ref{t1}. Following
the procedures given in
Ref.\cite{pentaquark1,pentaquark2,Ke:2007tg,Ke:2012wa} the
transition matrix element can be computed with the wavefunctions
of $\mid \Xi_{cc}(P,S,S_z) \ra$ and $\mid \Xi_{c}^(P',S',S'_z)
\ra$. The $cu$ subsystem stands as a spectator, i.e. its
spin configuration does not change during the transition, so the
transition matrix element can be
divided into two parts:
\begin{eqnarray}\label{s00}
&& \la \Xi_{c}(P',S_z') \mid \bar{s} \gamma^{\mu} (1-\gamma_{5}) c
\mid \Xi_{cc}(P,S_z) \ra = \frac{\sqrt{6}}{4}\la \Xi_{c}(P',S_z')
\mid \bar{s} \gamma^{\mu} (1-\gamma_{5}) c \mid \Xi_{cc}(P,S_z)
\ra_0\nonumber\\&&+\frac{\sqrt{6}}{4}\la \Xi_{c}(P',S_z') \mid
\bar{c} \gamma^{\mu} (1-\gamma_{5}) b \mid \Xi_{cc}(P,S_z) \ra_1
\end{eqnarray}
with
\begin{eqnarray}\label{s00}
&& \la \Xi_{c}(P',S_z') \mid \bar{s}
\gamma^{\mu} (1-\gamma_{5}) c \mid \Xi_{cc}(P,S_z) \ra_0  \nonumber \\
 &=& \int\frac{\{d^3 \tilde p_2\}\{d^3 \tilde p_3\}\phi_{\Xi_{c}}^*(x',k'_{\perp})
  \phi_{\Xi_{cc}}(x,k_{\perp})Tr[(\bar{P'}\!\!\!\!\!\slash'-M_0')\gamma_{5}(p_2\!\!\!\!\!\slash+m_2)(\bar{P}\!\!\!\!\!\slash'+M_0)\gamma_{5}(p_3\!\!\!\!\!\slash-m_3)]}{16\sqrt{p^+_1p'^+_1\bar{P}^+\bar{P'}^+M_0^3(m_1+e_1)
 (m_2+e_2)(m_3+e_3)(m_1'+e_1')
 (m_2'+e_2')(m_3'+e_3')}}\nonumber \\
  &&\times  \bar{u}(\bar{P'},S'_z)
  (p_1\!\!\!\!\!\slash'+m'_1)\gamma^{\mu}(1-\gamma_{5})
  (p_1\!\!\!\!\!\slash+m_1)  u(\bar{P},S_z),
\end{eqnarray}
and
\begin{eqnarray}\label{s01}
&& \la \Xi_{c}(P',S_z') \mid \bar{s}
\gamma^{\mu} (1-\gamma_{5}) c \mid \Xi_{cc}(P,S_z) \ra_1  \nonumber \\
 &=& \frac{\int\{d^3 \tilde p_2\}\{d^3 \tilde p_3\}\phi_{\Xi_{c}}^*(x',k'_{\perp})
  \phi_{\Xi_{cc}}(x,k_{\perp})Tr[\gamma_{\perp}^\alpha(\bar{P'}\!\!\!\!\!\slash'+M_0')\gamma_{5}(p_2\!\!\!\!\!\slash+m_2)(\bar{P}\!\!\!\!\!\slash'+M_0)\gamma_{5}\gamma_{\perp}^\beta(p_3\!\!\!\!\!\slash-m_3)]}{48\sqrt{p^+_1p'^+_1\bar{P}^+\bar{P'}^+M_0^3(m_1+e_1)
 (m_2+e_2)(m_3+e_3)(m_1'+e_1')
 (m_2'+e_2')(m_3'+e_3')}}\nonumber \\
  &&\times  \bar{u}(\bar{P'},S'_z)\gamma_{\perp\alpha}\gamma_{5}
  (p_1\!\!\!\!\!\slash'+m'_1)\gamma^{\mu}(1-\gamma_{5})
  (p_1\!\!\!\!\!\slash+m_1)  \gamma_{\perp\beta}\gamma_{5}u(\bar{P},S_z),
\end{eqnarray}
where
 \beq
m_1=m_c, \qquad m'_1=m_d, \qquad m_2=m_{c}, \qquad m_3=m_{u},
\qquad \gamma_{\perp\beta}=\gamma_{\beta}-v\!\!\!\slash v_{\beta}
 \eeq
and
$P$ ($P'$) is the four-momentum of $\Xi_{cc}$ ($\Xi_{c}$).
Setting $\tilde{p}_1=\tilde{p}'_1+\tilde{q}$, $\tilde{p}_2=\tilde{p}'_2$ and $\tilde{p}_3=\tilde{p}'_3$ we have
 \beq
 x_{1,2,3}'=x_{1,2,3}, \quad
 k'_{1\perp}=k_{1\perp}-(1-x_1)q_{\perp}, \quad
 k'_{2\perp}=k_{2\perp}+x_2q_{\perp}, \quad
 k'_{3\perp}=k_{3\perp}+x_3q_{\perp}.
 \eeq

The form factors for the weak transition $\Xi_{cc}\rightarrow
\Xi_{c}$  are defined in the standard way as
\begin{eqnarray}\label{s1}
&& \la \Xi_{c}(P',S',S_z') \mid \bar{s}\gamma_{\mu}
 (1-\gamma_{5})c \mid \Xi_{cc}(P,S,S_z) \ra  \non \\
 &=& \bar{u}_{\Xi_{c}}(P',S'_z) \left[ \gamma_{\mu} f_{1}(q^{2})
 +i\sigma_{\mu \nu} \frac{ q^{\nu}}{M_{\Xi_{cc}}}f_{2}(q^{2})
 +\frac{q_{\mu}}{M_{\Xi_{cc}}} f_{3}(q^{2})
 \right] u_{\Xi_{cc}}(P,S_z) \nonumber \\
 &&-\bar u_{\Xi_{c}}(P',S'_z)\left[\gamma_{\mu} g_{1}(q^{2})
  +i\sigma_{\mu \nu} \frac{ q^{\nu}}{M_{\Xi_{c}}}g_{2}(q^{2})+
  \frac{q_{\mu}}{M_{\Xi_{cc}}}g_{3}(q^{2})
 \right]\gamma_{5} u_{\Xi_{cc}}(P,S_z).
\end{eqnarray}
where  $q \equiv P-P'$. For $\la \Xi_{c}(P',S',S_z')
\mid \bar{s}\gamma_{\mu}
 (1-\gamma_{5})c \mid \Xi_{cc}(P,S,S_z) \ra_0$ and $\la \Xi_{c}(P',S',S_z') \mid
\bar{s}\gamma_{\mu}
 (1-\gamma_{5})c \mid \Xi_{cc}(P,S,S_z) \ra_1$ the form factors are denoted to $f_i^s$, $g_i^s$  and $f_i^v$ , $g_i^v$, so we have
\begin{eqnarray}\label{relation}
f_1=\frac{\sqrt{6}}{4}f^s_1+\frac{\sqrt{6}}{4}f^v_1,
g_1=\frac{\sqrt{6}}{4}g^s_1+\frac{\sqrt{6}}{4}g^v_1,\nonumber\\
f_2=\frac{\sqrt{6}}{4}f^s_2+\frac{\sqrt{6}}{4}f^v_2,
g_2=\frac{\sqrt{6}}{4}g^s_2+\frac{\sqrt{6}}{4}g^v_2.
\end{eqnarray}\label{relation}

In our earlier paper\cite{Ke:2019smy} $f_i^s$, $g_i^s$, $f_i^v$ and
$g_i^v$ are presented as
\begin{eqnarray}\label{s21}
f^s_1
 &=& \int\frac{ d x_2 d^2 k^2_{2\perp}}{2(2\pi)^3}\frac{ d x_3 d^2 k^2_{3\perp}}{2(2\pi)^3}
 \frac{{\rm Tr}[(\bar{P'}\!\!\!\!\!\slash-M_0')\gamma_{5}(p_2\!\!\!\!\!\slash+m_2)(\bar{P}\!\!\!\!\!\slash+M_0)
 \gamma_{5}(p_3\!\!\!\!\!\slash-m_3)]}{\sqrt{M_0^3(m_1+e_1)
 (m_2+e_2)(m_3+e_3)(m_1'+e_1')
 (m_2'+e_2')(m_3'+e_3')}}\nonumber \\
  &&\times\frac{\phi_{\Xi_{c}}^*(x',k'_{\perp})
  \phi_{\Xi_{cc}}(x,k_{\perp})}{16\sqrt{x_1x'_1}} \frac{{\rm Tr}[ (\bar{P}\!\!\!\!\!\slash+M_0)\gamma^+(\bar{P'}\!\!\!\!\!\slash+M_0')
  (p_1\!\!\!\!\!\slash'+m'_1)\gamma^{+}
  (p_1\!\!\!\!\!\slash+m_1)  ]}{8P^+P'^+},
\nonumber\\
\frac{f^s_2}{M_{\Xi_{cc}}}
 &=& \frac{-i}{q_{\perp}^i}\int\frac{ d x_2 d^2 k^2_{2\perp}}{2(2\pi)^3}\frac{ d x_3 d^2 k^2_{3\perp}}{2(2\pi)^3}
 \frac{{\rm Tr}[(\bar{P'}\!\!\!\!\!\slash-M_0')\gamma_{5}(p_2\!\!\!\!\!\slash+m_2)(\bar{P}\!\!\!\!\!\slash+M_0)\gamma_{5}(p_3\!\!\!\!\!\slash-m_3)]}{\sqrt{M_0^3(m_1+e_1)
 (m_2+e_2)(m_3+e_3)(m_1'+e_1')
 (m_2'+e_2')(m_3'+e_3')}}\nonumber \\
  &&\times\frac{\phi_{\Xi_{c}}^*(x',k'_{\perp})
  \phi_{\Xi_{cc}}(x,k_{\perp})}{16\sqrt{x_1x'_1}} \frac{{\rm Tr}[ (\bar{P}\!\!\!\!\!\slash+M_0)\sigma^{i+}(\bar{P'}\!\!\!\!\!\slash+M_0')
  (p_1\!\!\!\!\!\slash'+m'_1)\gamma^+
  (p_1\!\!\!\!\!\slash+m_1)  ]}{8P^+P'^+},
\nonumber\\
g^s_1
 &=& \int\frac{ d x_2 d^2 k^2_{2\perp}}{2(2\pi)^3}\frac{ d x_3 d^2 k^2_{3\perp}}{2(2\pi)^3}
 \frac{{\rm Tr}[(\bar{P'}\!\!\!\!\!\slash-M_0')\gamma_{5}(p_2\!\!\!\!\!\slash+m_2)(\bar{P}\!\!\!\!\!\slash+M_0)
 \gamma_{5}(p_3\!\!\!\!\!\slash-m_3)]}{\sqrt{M_0^3(m_1+e_1)
 (m_2+e_2)(m_3+e_3)(m_1'+e_1')
 (m_2'+e_2')(m_3'+e_3')}}\nonumber \\
  &&\times\frac{\phi_{\Xi_{c}}^*(x',k'_{\perp})
  \phi_{\Xi_{cc}}(x,k_{\perp})}{16\sqrt{x_1x'_1}} \frac{{\rm Tr}[ (\bar{P}\!\!\!\!\!\slash+M_0)\gamma^+\gamma_5(\bar{P'}\!\!\!\!\!\slash+M_0')
  (p_1\!\!\!\!\!\slash'+m'_1)\gamma^{+}\gamma_5
  (p_1\!\!\!\!\!\slash+m_1)  ]}{8P^+P'^+},
\nonumber\\
\frac{g^s_2}{M_{\Xi_{cc}}}
 &=& \frac{i}{q_{\perp}^i}\int\frac{ d x_2 d^2 k^2_{2\perp}}{2(2\pi)^3}\frac{ d x_3 d^2 k^2_{3\perp}}{2(2\pi)^3}
 \frac{{\rm Tr}[(\bar{P'}\!\!\!\!\!\slash-M_0')\gamma_{5}(p_2\!\!\!\!\!\slash+m_2)(\bar{P}\!\!\!\!\!\slash+M_0)
 \gamma_{5}(p_3\!\!\!\!\!\slash-m_3)]}{\sqrt{M_0^3(m_1+e_1)
 (m_2+e_2)(m_3+e_3)(m_1'+e_1')
 (m_2'+e_2')(m_3'+e_3')}}\nonumber \\
  &&\times\frac{\phi_{\Xi_{c}}^*(x',k'_{\perp})
  \phi_{\Xi_{cc}}(x,k_{\perp})}{16\sqrt{x_1x'_1}} \frac{{\rm Tr}[ (\bar{P}\!\!\!\!\!\slash+M_0)\sigma^{i+}\gamma_5(\bar{P'}\!\!\!\!\!\slash+M_0')
  (p_1\!\!\!\!\!\slash'+m'_1)\gamma^+\gamma_5
  (p_1\!\!\!\!\!\slash+m_1)  ]}{8P^+P'^+},
\nonumber\\
f^v_1
 &=&\int\frac{ d x_2 d^2 k^2_{2\perp}}{2(2\pi)^3}\frac{ d x_3 d^2 k^2_{3\perp}}{2(2\pi)^3}
 \frac{{\rm Tr}[\gamma_{\perp}^\alpha(\bar{P'}\!\!\!\!\!\slash'+M_0')\gamma_{5}(p_2\!\!\!\!\!\slash+m_2)(\bar{P}\!\!\!\!\!\slash'+M_0)
 \gamma_{5}\gamma_{\perp}^\beta(p_3\!\!\!\!\!\slash-m_3)]]}{\sqrt{M_0^3(m_1+e_1)
 (m_2+e_2)(m_3+e_3)(m_1'+e_1')
 (m_2'+e_2')(m_3'+e_3')}}\nonumber \\
  &&\times\frac{\phi_{\Xi_{c}}^*(x',k'_{\perp})
  \phi_{\Xi_{cc}}(x,k_{\perp})}{48\sqrt{x_1x'_1}}  \frac{{\rm Tr}[ (\bar{P}\!\!\!\!\!\slash+M_0)\gamma^+(\bar{P'}\!\!\!\!\!\slash+M_0')
  \gamma_{\perp\alpha}\gamma_{5}(p_1\!\!\!\!\!\slash'+m'_1)\gamma^{+}
  (p_1\!\!\!\!\!\slash+m_1)\gamma_{\perp\beta}\gamma_{5}  ]}{8P^+P'^+},
\nonumber\\
\frac{f^v_2}{M_{\Xi_{cc}}}
 &=& \frac{-i}{q_{\perp}^i}\int\frac{ d x_2 d^2 k^2_{2\perp}}{2(2\pi)^3}\frac{ d x_3 d^2 k^2_{3\perp}}{2(2\pi)^3}
 \frac{{\rm Tr}[\gamma_{\perp}^\alpha(\bar{P'}\!\!\!\!\!\slash'+M_0')\gamma_{5}(p_2\!\!\!\!\!\slash+m_2)(\bar{P}\!\!\!\!\!\slash'+M_0)\gamma_{5}
 \gamma_{\perp}^\beta(p_3\!\!\!\!\!\slash-m_3)]}{\sqrt{M_0^3(m_1+e_1)
 (m_2+e_2)(m_3+e_3)(m_1'+e_1')
 (m_2'+e_2')(m_3'+e_3')}}\nonumber \\
  &&\times\frac{\phi_{\Xi_{c}}^*(x',k'_{\perp})
  \phi_{\Xi_{cc}}(x,k_{\perp})}{48\sqrt{x_1x'_1}} \frac{{\rm Tr}[ (\bar{P}\!\!\!\!\!\slash-M_0)\sigma^{i+}(\bar{P'}\!\!\!\!\!\slash-M_0')
\gamma_{\perp\alpha}\gamma_{5}(p_1\!\!\!\!\!\slash'+m'_1)\gamma^{+}
  (p_1\!\!\!\!\!\slash+m_1)\gamma_{\perp\beta}\gamma_{5}   ]}{8P^+P'^+},
\nonumber\\
g^v_1
 &=& \int\frac{ d x_2 d^2 k^2_{2\perp}}{2(2\pi)^3}\frac{ d x_3 d^2 k^2_{3\perp}}{2(2\pi)^3}
 \frac{{\rm Tr}[\gamma_{\perp}^\alpha(\bar{P'}\!\!\!\!\!\slash'+M_0')\gamma_{5}(p_2\!\!\!\!\!\slash+m_2)(\bar{P}\!\!\!\!\!\slash'+M_0)
 \gamma_{5}\gamma_{\perp}^\beta(p_3\!\!\!\!\!\slash-m_3)]}{\sqrt{M_0^3(m_1+e_1)
 (m_2+e_2)(m_3+e_3)(m_1'+e_1')
 (m_2'+e_2')(m_3'+e_3')}}\nonumber \\
  &&\times\frac{\phi_{\Xi_{c}}^*(x',k'_{\perp})
  \phi_{\Xi_{cc}}(x,k_{\perp})}{48\sqrt{x_1x'_1}} \frac{{\rm Tr}[ (\bar{P}\!\!\!\!\!\slash-M_0)\gamma^+\gamma_5(\bar{P'}\!\!\!\!\!\slash-M_0')
  \gamma_{\perp\alpha}\gamma_{5}(p_1\!\!\!\!\!\slash'+m'_1)\gamma^{+}
  (p_1\!\!\!\!\!\slash+m_1)\gamma_{\perp\beta}\gamma_{5}  ]}{8P^+P'^+},
\nonumber\\
\frac{g^v_2}{M_{\Xi_{cc}}}
 &=& \frac{i}{q_{\perp}^i}\int\frac{ d x_2 d^2 k^2_{2\perp}}{2(2\pi)^3}\frac{ d x_3 d^2 k^2_{3\perp}}{2(2\pi)^3}
 \frac{{\rm Tr}[\gamma_{\perp}^\alpha(\bar{P'}\!\!\!\!\!\slash'+M_0')\gamma_{5}(p_2\!\!\!\!\!\slash+m_2)(\bar{P}\!\!\!\!\!\slash'+M_0)\gamma_{5}
\gamma_{\perp}^\beta(p_3\!\!\!\!\!\slash-m_3)]}{\sqrt{M_0^3(m_1+e_1)
 (m_2+e_2)(m_3+e_3)(m_1'+e_1')
 (m_2'+e_2')(m_3'+e_3')}}\nonumber \\
  &&\times\frac{\phi_{\Xi_{c}}^*(x',k'_{\perp})
  \phi_{\Xi_{cc}}(x,k_{\perp})}{48\sqrt{x_1x'_1}} \frac{{\rm Tr}[ (\bar{P}\!\!\!\!\!\slash-M_0)\sigma^{i+}\gamma_5(\bar{P'}\!\!\!\!\!\slash-M_0')
  \gamma_{\perp\alpha}\gamma_{5}(p_1\!\!\!\!\!\slash'+m'_1)\gamma^{+}
  (p_1\!\!\!\!\!\slash+m_1)\gamma_{\perp\beta}\gamma_{5}  ]}{8P^+P'^+}.
\end{eqnarray}

For the transition $\la \Xi'_{c}(P',S',S_z') \mid \bar{Q}'\gamma_{\mu}
 (1-\gamma_{5})Q \mid \Xi_{cc}(P,S,S_z) \ra$
the form factors are also defined as in Eq. (\ref{s1}). Here we
just add `` $'$ " on $f_1$, $f_2$, $g_1$ and $g_2$ in order to
distinguish the quantities for $\Xi_{cc}\to\Xi_c$ and those for
$\Xi_{cc}\to \Xi'_c$. They are
\begin{eqnarray}\label{relation}
f'_1=-\frac{3\sqrt{2}}{4}f^s_1+\frac{\sqrt{2}}{4}f^v_1,
g'_1=-\frac{3\sqrt{2}}{4}g^s_1+\frac{\sqrt{2}}{4}g^v_1,\nonumber\\
f'_2=-\frac{3\sqrt{2}}{4}f^s_2+\frac{\sqrt{2}}{4}f^v_2,
g'_2=-\frac{3\sqrt{2}}{4}g^s_2+\frac{\sqrt{2}}{4}g^v_2.
\end{eqnarray}\label{relation}
In the calculation one also needs to use
$\phi_{\Xi'_{c}}^*(x',k'_{\perp})$ to replace
$\phi_{\Xi_{c}}^*(x',k'_{\perp})$.

\section{Numerical Results}

\begin{table}
\caption{The quark mass and the parameter $\beta$ (in units of
 GeV).}\label{Tab:t1}
\begin{ruledtabular}
\begin{tabular}{cccccc}
  $m_c$  & $m_s$  &$m_{u}$ & $\beta_{c[cu]}$ & $\beta_{s[cu]}$& $\beta_{[cu]}$ \\\hline
  $1.5$  & $0.5$  & 0.25     & 1.898     &0.760  & 0.656
\end{tabular}
\end{ruledtabular}
\end{table}

\subsection{$\Xi_{cc}\to \Xi_c$ and $\Xi_{cc}\to \Xi'_c$ form factors }
Before we start to evaluate those form factors numerically the parameters
in the concerned model are needed to be determined. The masses of
quarks given in Ref.\cite{Chang:2018zjq} are collected
in table I. The masses of $\Xi_{c}$ and $\Xi'_{c}$ are taken from
\cite{PDG18}. Indeed, we know very little about the parameters
$\beta_1$ and $\beta_{23}$ in the wave function of the initial
baryon and $\beta_1'$ and $\beta_{23}'$ in that of the final
baryon. Generally the reciprocal of $\beta$ is related to the
electrical radium of two constituents. Since the strong
interaction between $q$ and $q^{(')}$ is a half of that between
$q\bar q^{(')}$, if it is a Coulomb-like potential one can expect
the the electrical radium of $qq^{(')}$ to be $1/\sqrt{2}$ times
that of $q\bar q^{(')}$ i.e.
$\beta_{qq^{(')}}\approx\sqrt{2}\beta_{q\bar q^{(')}}$. In
Ref.\cite{LeYaouanc:1988fx} considering the binding energy the
authors obtained the same results. In our early paper
for a compact $q q^{(')}$ system we find
$\beta_{qq^{(')}}=2.9\beta_{q\bar q^{(')}}$ i.e   the electrical
radium of $qq^{(')}$ to be $1/2.9$ times that of $q\bar q^{(')}$.
In terms of the knowledge we can estimate $\beta_{c[cu]}\approx
2.9\beta_{c\bar c}$, $\beta_{s[cu]}\approx \sqrt{2}\beta_{c\bar
s}$, $\beta_{[cu]}\approx \sqrt{2}\beta_{c\bar u}$ where
$\beta_{c\bar c}$, $\beta_{c\bar u}$ and $\beta_{c\bar s}$ were
obtained for the mesons case\cite{Chang:2018zjq}. With these
parameters we calculate the form factors and make theoretical
predictions on the transition rates.

\begin{figure}
\begin{center}
\scalebox{0.8}{\includegraphics{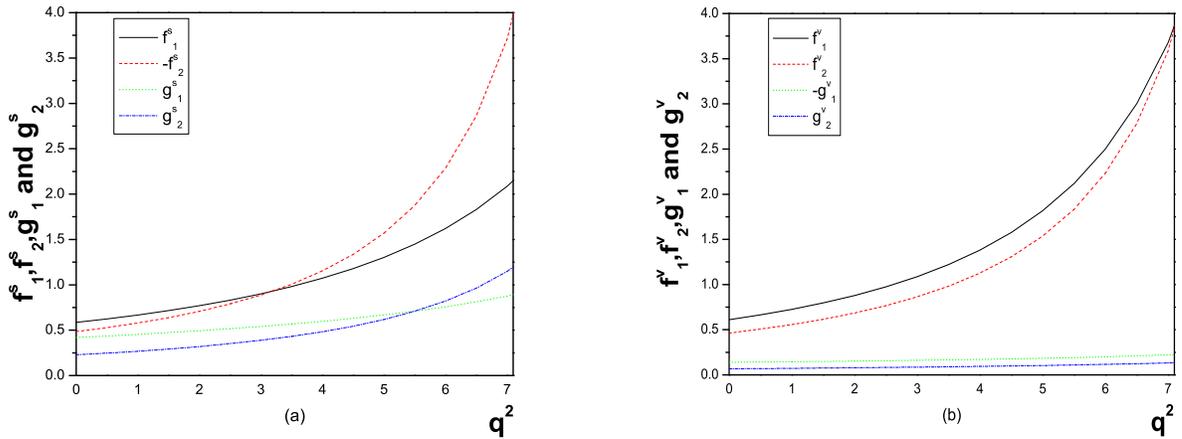}}
\end{center}
\caption{The dependence of  form factors $f^s_1$, $f^s_2$, $g^s_1$ and $g^s_2$
in a three-parameter form on $q^2$ (a) and The dependence of the form factors
$f^v_1$, $f^v_2$, $g^v_1$ and $g^v_2$ on $q^2$ (b)
.}\label{fg1}
\end{figure}

 Since these  form
factors $f^{s(v)}_i\,{(i=1,2)}$ and $g^{s(v)}_i\,{(i=1,2)}$ are
evaluated in the frame $q^+=0$ i.e. $q^2=-q^2_{\perp}\leq 0$ (the
space-like region) one needs to extend them into the time-like
region. In Ref.\cite{pentaquark2} a three-parameter form was employed
 \begin{eqnarray}\label{s145}
 F(q^2)=\frac{F(0)}{\left(1-\frac{q^2}{M_{\Xi_{cc}}^2}\right)
  \left[1-a\left(\frac{q^2}{M_{\Xi_{cc}}^2}\right)
  +b\left(\frac{q^2}{M_{\Xi_{cc}}^2}\right)^2\right]},
 \end{eqnarray}
where $F(q^2)$ denotes the form factors $f^{s(v)}_i$ and
$g^{s(v)}_i$.
 Using the form factors in the space-like region
we may calculate numerically the parameters $a,~b$ and $F(0)$ in
the un-physical region, namely fix $F(q^2\leq 0)$. As discussed in
previous section, these forms are extended into the physical
region with $q^2\geq 0$ through Eq. (\ref{s145}). The fitted values of
$a,~b$ and $F(0)$ in the form factors $f_{1}$, $f_{1}$, $g_{1}$
and $g_{2}$ are presented in Table \ref{Tab:t1}. The dependence of
the form factors on $q^2$ is depicted in Fig. \ref{fg1}.
\begin{table}
\caption{The form factors given in the
  three-parameter form.}\label{Tab:t1}
\begin{ruledtabular}
\begin{tabular}{cccc}
  $F$    &  $F(0)$ &  $a$  &  $b$ \\\hline
  $f^s_1$  &   0.586    & 0.640    & -0.194   \\
$f^s_2$  &   -0.484     & 1.23   &-0.222   \\
  $g^s_1$  &      0.420    &     -0.0142&  0.0748  \\
  $g^s_2$  &      0.228  &    1.02  &  -0.101  \\
  $f^v_1$  &   0.610     &  1.18    & -0.0492   \\
$f^v_2$  &   0.463     &   1.32    & -0.0642  \\
  $g^v_1$  &      -0.140    &    -0.501 & 0.274 \\
  $g^v_2$  &      0.0673   &    0.00936  &  0.327
\end{tabular}
\end{ruledtabular}
\end{table}

\begin{figure}
\begin{center}
\scalebox{0.8}{\includegraphics{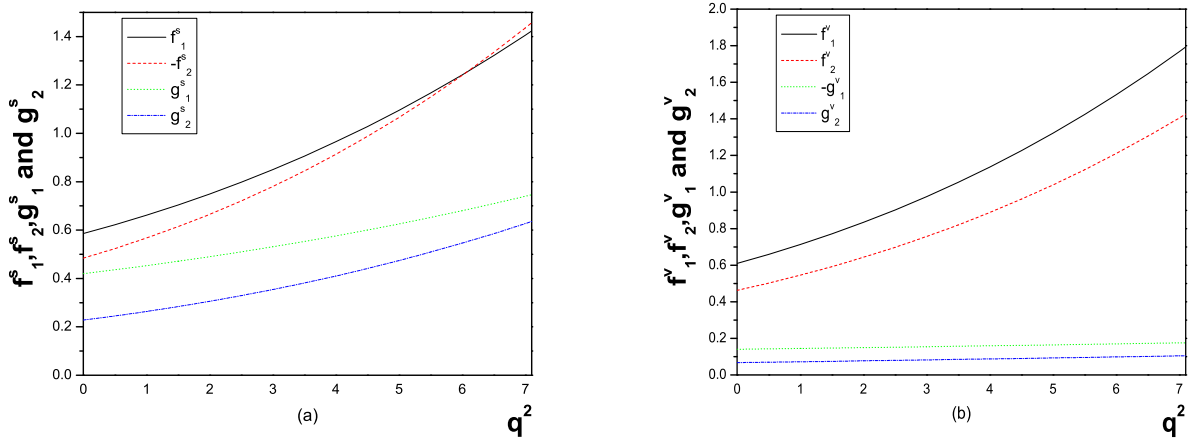}}
\end{center}
\caption{The dependence of  form factors $f^s_1$, $f^s_2$, $g^s_1$ and $g^s_2$
in polynomial form on $q^2$ (a) and The dependence of  form factors
$f^v_1$, $f^v_2$, $g^v_1$ and $g^v_2$ on $q^2$ (b)
.}\label{fg2}
\end{figure}

Since the form factor $f^s_1$, $f^s_1$, $f^v_1$ and $f^v_2$
rise quickly after $q^2>6$ GeV which are very different with the
results in other $\frac{1}{2}\to \frac{1}{2} $
transitions\cite{pentaquark2,Ebert:2006rp,Korner:1992wi,Ke:2007tg,Ke:2019smy},
we suggest a polynomial to parameterize these form factors
 \begin{eqnarray}\label{s146}
 F(q^2)=F(0)\left[1+a'\left(\frac{q^2}{M_{\Xi_{cc}}^2}\right)
  +b'\left(\frac{q^2}{M_{\Xi_{cc}}^2}\right)^2+c'\left(\frac{q^2}{M_{\Xi_{cc}}^2}\right)^3\right].
 \end{eqnarray}
The fitted values of $a',~b',~c'$ and $F(0)$ in the form factors
are presented in Table \ref{Tab:t2}. The dependence of the form
factors on $q^2$ is depicted in Fig. \ref{fg2}. The
figures of these form factors are apparently more smooth and similar to those for
$\frac{1}{2}\to\frac{1}{2}$ transition
\cite{pentaquark2,Ebert:2006rp,Korner:1992wi,Ke:2007tg,Ke:2019smy}.

\begin{table}
\caption{The form factors given in the
 ploynomial form.}\label{Tab:t2}
\begin{ruledtabular}
\begin{tabular}{ccccc}
  $F$    &  $F(0)$ &  $a'$  &  $b'$ & $c'$\\\hline
  $f^s_1$  &   0.586    & 1.57    &1.59 & 0.704   \\
$f^s_2$  &   -0.484     &2.06  &2.42 &1.17   \\
  $g^s_1$  &      0.420    &    0.983& 0.692& 0.258  \\
  $g^s_2$  &      0.228  &    1.90  &  2.07 &0.960  \\
  $f^v_1$  &   0.610     &  2.04    &2.27 & 1.06   \\
$f^v_2$  &   0.463     &  2.14    & 2.49 & 1.19  \\
  $g^v_1$  &      -0.140    &    0.422 & 0.0931 & 0.00632 \\
  $g^v_2$  &      0.0673   &    0.925  &  0.245 & -0.0862
\end{tabular}
\end{ruledtabular}
\end{table}

\subsection{Semi-leptonic decay of $\Xi_{cc} \to
\Xi_{c} +l\bar{\nu}_l$ and $\Xi_{cc} \to \Xi'_{c} +l\bar{\nu}_l$}

\begin{figure}[hhh]
\begin{center}
\scalebox{0.8}{\includegraphics{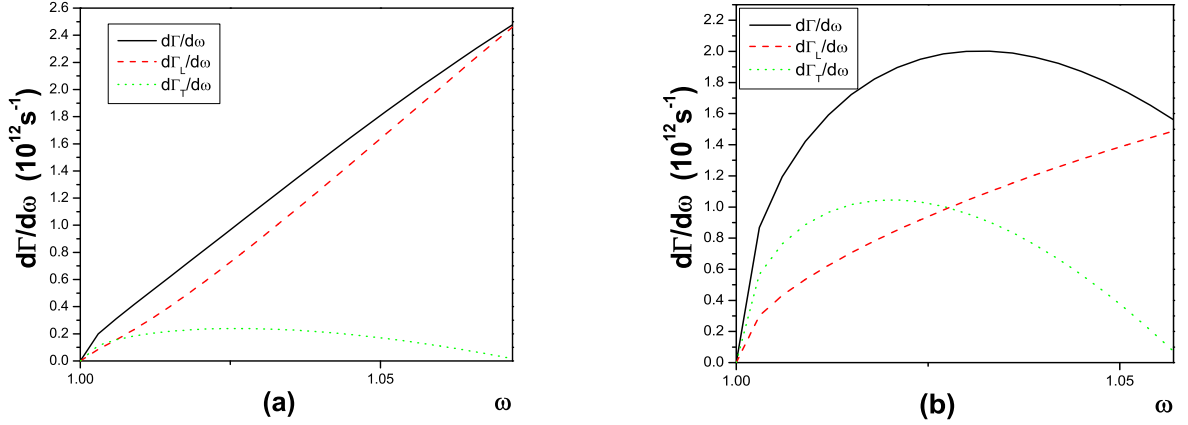}}
\end{center}
\caption{ Differential decay rates $d\Gamma/d\omega$ for the decay
$\Xi_{cc} \to \Xi_{c} l\bar{\nu}_l$(a) and $\Xi'_{cc} \to \Xi_{c}
l\bar{\nu}_l$ (b)}\label{f3}
\end{figure}

Using the form factors obtained in last subsection, we
evaluate the rate of $\Xi_{cc} \to \Xi_{c} l\bar{\nu}_l$ and
$\Xi_{cc} \to \Xi'_{c} l\bar{\nu}_l$. The differential decay
widths $d\Gamma/d\omega$ ($\omega=v\cdot v'$) are depicted in
Fig. \ref{f3}. Our predictions on the total decay widths,
longitudinal decay widths, transverse decay widths and the ratio
of the longitudinal to transverse decay rates $R$ are also listed in table
\ref{Tab:t4}. Deliberately letting the quark masses and all $\beta$s
fluctuate up to 5\% , one can estimate possible theoretical uncertainties of the numerical
results.

In Ref.\cite{Wang:2017mqp}  the authors employ three-parameter
parametrization scheme to fix theses form factor and their
predictions on $\Gamma(\Xi_{cc} \to \Xi_{c} l\bar{\nu}_l)$ and
$\Gamma(\Xi'_{cc} \to \Xi_{c}) l\bar{\nu}_l$ are almost twice
larger than our results presented in table \ref{Tab:t4}. Our
estimate on the ratio of longitudinal to transverse decay rates
$R$ for $\Xi'_{cc} \to \Xi_{c} l\bar{\nu}_l$ is close to 1.42
given in \cite{Wang:2017mqp} but that of $\Xi_{cc} \to \Xi_{c}
l\bar{\nu}_l$ slightly deviates from theirs. One also notices that
the predictions on $\Gamma(\Xi_{cc} \to \Xi_{c} l\bar{\nu}_l)$ are
close to each others in different approaches (not include the
prediction in \cite{Wang:2017mqp} ) but those on $\Gamma(\Xi_{cc}
\to \Xi'_{c} l\bar{\nu}_l)$ deviate from each others a little bit.
From the values in table \ref{Tab:t4} one can find our
$\Gamma(\Xi_{cc} \to \Xi_{c} l\bar{\nu}_l)$ is very close to
$\Gamma(\Xi'_{cc} \to \Xi_{c} l\bar{\nu}_l)$ which is consistent
with the prediction under SU(3) limit.

\begin{table}
\caption{The width (in unit $10^{12}{\rm s}^{-1}$)   of
$\Xi_{cc}\to \Xi_{c} l\bar{\nu}_l$ (left) and $\Xi_{cc}\to
\Xi'_{c} l\bar{\nu}_l$ (right).}\label{Tab:t4}
\begin{ruledtabular}
\begin{tabular}{ccc|ccc}
  &  $\Gamma$ &  R  &  $\Gamma_T$ & $R$    \\\hline
 this work &  0.100$\pm$0.015  & 7.14$\pm$0.61& 0.0995$\pm0.0091$  &  1.34$\pm$0.07\\\hline
 Ref.\cite{Wang:2017mqp} &0.173  &9.99 &0.193 &1.42\\\hline
Ref.\cite{Shi:2019hbf} &$0.092\pm0.014$  &$22\pm8$
&$0.032\pm0.006$ &$1.1\pm0.2$\\\hline Ref.\cite{Gutsche:2019iac}
&0.106 &- &$0.147$ &-
\end{tabular}
\end{ruledtabular}
\end{table}

\subsection{Non-leptonic decays of ${\Xi_{cc}}\to\Xi_{c}+ M$ and ${\Xi_{cc}}\to\Xi'_{c}+ M$}

On the theoretical aspect, calculating the concerned quantities
of the non-leptonic decays seems to be more complicated than for
semi-leptonic processes.
Our theoretical framework is based on the
factorization assumption, namely the hadronic transition matrix
element is factorized into a product of two independent hadronic matrix
elements of currents,
\begin{eqnarray}\label{s0}
&& \la \Xi^{(')}_{c}(P',S_z')M \mid \mathcal{H} \mid \Xi_{cc}(P,S_z) \ra  \nonumber \\
 &=&\frac{G_FV_{cs}V^*_{qq'}}{\sqrt{2}}\la M \mid
\bar{q'} \gamma^{\mu} (1-\gamma_{5}) q \mid 0\ra\la
\Xi^{(')}_{c}(P',S_z') \mid \bar{s} \gamma^{\mu} (1-\gamma_{5})c
\mid \Xi_{cc}(P,S_z) \ra,
\end{eqnarray}
where the term $\la M \mid \bar{q'} \gamma^{\mu} (1-\gamma_{5}) q
\mid 0\ra$ is determined by a decay constant and the transition
$\Xi_{cc}\rightarrow \Xi^{(')}_{c}$ is evaluated in the previous
sections.
Since the decays $\Xi_{cc}\to\Xi^{(')}_{c}+M$ is the so-called
color-favored portal, the factorization should be a plausible
approximation. The results on these non-leptonic decays can be checked
in the coming measurements and
the validity degree of the obtained form factors in the doubly
charmed baryon would be further examined.

From the results shown in Tab. \ref{Tab:t6}, we find
$\Xi_{cc}\to\Xi^{(')}_{c} \pi$  and $\Xi_{cc}\to\Xi^{(')}_{c}
\rho$ are the main two-body decay channels for $\Xi_{cc}$.
Especially $\Gamma(\Xi_{cc}\to\Xi^{'}_{c} \rho)$ is close to
$\Gamma(\Xi_{cc}\to\Xi_{c} \pi)$ which should also be observed in
LHCb soon. The predictions in other approaches are listed in Tab.
\ref{Tab:t61}. The theoretical predictions on the widths
calculated in Ref.\cite{Wang:2017mqp} are two or three times
larger than ours. The results on $\Gamma(\Xi_{cc}\to\Xi^{'}_{c}
\pi)$ and $\Gamma(\Xi_{cc}\to\Xi^{'}_{c} K)$  in
Ref.\cite{Shi:2019hbf} are close to ours but there exists still a
discrepancy for other channels. This should also be tested in the
future more precise measurements.

\begin{table}
\caption{Our predictions on Widths (in unit $10^{10}{\rm s}^{-1}$)
and up-down asymmetry of non-leptonic decays
$\Xi_{cc}\to\Xi^{(')}_{c} M$.}\label{Tab:t6}
\begin{ruledtabular}
\begin{tabular}{ccc|ccc}
 mode& width & up-down asymmetry &mode &width  &up-down asymmetry\\\hline
  $\Xi_{cc}\to\Xi_{c} \pi$ & 13.6$\pm$1.8&-0.441$\pm$0.009&$\Xi_{cc}\to\Xi'_{c} \pi$  &7.68$\pm$0.92 &-0.982$\pm$0.005 \\\hline
  $\Xi_{cc}\to\Xi_{c} \rho$ &11.0$\pm$1.5&-0.429$\pm$0.016&
  $\Xi_{cc}\to\Xi'_{c} \rho$&13.9$\pm$1.2&-0.111$\pm$0.034\\\hline
  $\Xi_{cc}\to\Xi_{c} K$   &1.03$\pm$0.14&-0.402$\pm$0.008&$\Xi_{cc}\to\Xi'_{c}
  K$&0.492$\pm$0.059&-0.998$\pm$0.002
                            \\\hline
  $\Xi_{cc}\to\Xi_{c} K^{*}$ &
  0.414$\pm$0.055&-0.422$\pm$0.021&
  $\Xi_{cc}\to\Xi'_{c} K^{*}$&0.623$\pm$0.052&-0.014$\pm$0.030
\end{tabular}
\end{ruledtabular}
\end{table}

\begin{table}
\caption{Widths (in unit $10^{10}{\rm s}^{-1}$) of non-leptonic
decays $\Xi_{cc}\to\Xi^{(')}_{c} M$ in references.}\label{Tab:t61}
\begin{ruledtabular}
\begin{tabular}{ccc|cccc}
 mode&\cite{Wang:2017mqp} &\cite{Shi:2019hbf} &  mode& \cite{Wang:2017mqp}&\cite{Shi:2019hbf}   &\cite{Gutsche:2019iac}\\\hline
  $\Xi_{cc}\to\Xi_{c} \pi$ &23.9&$12.0\pm1.7$ &$\Xi_{cc}\to\Xi'_{c} \pi$  &16.7 &$3.64\pm0.76$&11.9 \\\hline
  $\Xi_{cc}\to\Xi_{c} \rho$ &46.0&$24.3\pm3.1$&
  $\Xi_{cc}\to\Xi'_{c} \rho$&62.6&$9.72\pm1.98$ &62.9\\\hline
  $\Xi_{cc}\to\Xi_{c} K$   &1.99&$0.972\pm0.152$ &$\Xi_{cc}\to\Xi'_{c}
  K$&1.14&$0.334\pm0.061$&-
                            \\\hline
  $\Xi_{cc}\to\Xi_{c} K^{*}$ & 1.81&$0.972\pm0.152$&
  $\Xi_{cc}\to\Xi'_{c} K^{*}$&2.84&$0.349\pm0.05$&-
\end{tabular}
\end{ruledtabular}
\end{table}

\section{Conclusions and discussions}

In this paper we calculate the transition rate of
$\Xi_{cc}\to\Xi^{(')}_{c}$ in the light front quark model. For the
baryons $\Xi_{cc}$ and $\Xi^{(')}_{c}$ we employ a three-quark
picture instead of the quark-diquark one in our calculation.
Generally, two charm quarks constitute a diquark  and this widely
accepted scenario determines the wavefunction of $\Xi_{cc}$.
Because two $c$  quarks are  identical heavy flavor particles in a
color anti-triplet, it must be a vector-bosonic state, whereas, in
$\Xi^{(')}_{c}$ the light $us$ pair is seen as a diquark. In the
concerned process, the diquark in the initial state is different
from that in the final state, so that the diquark is no longer a
spectator and the diquark picture cannot be directly applied in
this case. However in the process the charm quark which does not
undergo a transition and the $u$ quark are approximatively
spectators when higher order QCD effects are neglected, so the
$cu$ pair can be regarded as an effective subsystem. Baryon is a
three-body-system whose total spin can be obtained through
different schemes just in analog to the L-S coupling and J-J
coupling in the quantum mechanics. Making a Racah transformation
we can convert one configuration into another. The  Racah
coefficients of such transformation determines correlation between
the two configuration $(cc)u$ and $(c(cu)$. However, one is noted
that the subsystem of $(cu)$ is not a diquark in a rigorous
meaning and the two constituents there exists a relative momentum.
Thus in the vertex function of the three-body system there exists
an inner degree of freedom for the $cu$ subsystem.

We calculate the form factors for the transitions
$\Xi_{cc}\to\Xi_{c}$ and $\Xi_{cc}\to\Xi^{'}_{c}$ in the
space-like region. When we extend them to physical region we find
three-parameter form is not a good choice, so that we suggest
to parameterize these form factors in terms of polynomials. Using these form
factors we calculate the rates of semileptonic decays
$\Xi_{cc}\to\Xi_{c}l\nu_l $ and $\Xi_{cc}\to\Xi^{'}_{c}l\nu_l $.
We find that $\Gamma(\Xi_{cc} \to \Xi_{c} l\bar{\nu}_l)$ is very close to
$\Gamma(\Xi_{cc} \to \Xi'_{c} l\bar{\nu}_l)$ and this conclusion is consistent
with the prediction under SU(3) limit but our results on
$\Gamma(\Xi_{cc} \to \Xi_{c} l\bar{\nu}_l)$ and $\Gamma(\Xi_{cc}
\to \Xi'_{c} l\bar{\nu}_l)$ are about a half of those in
Ref.\cite{Wang:2017mqp}. The ratio of the longitudinal to transverse
decay rates $R$ for $\Xi_{cc} \to \Xi_{c} l\bar{\nu}_l$ and
$\Xi_{cc} \to \Xi'_{c} l\bar{\nu}_l$ are roughly consistent with the
predictions of \cite{Wang:2017mqp} .
With the same theoretical framework, we also evaluate the rates of
several non-leptonic decays. Our numerical results indicate that the channel $\Xi_{cc} \to
\Xi_{c} \pi $ has the largest branching ratio for the transition
$\Xi_{cc} \to \Xi_{c}$,  instead, the channel $\Xi_{cc} \to \Xi'_{c} \rho $ is
the main channel for the transition $\Xi_{cc} \to
\Xi'_{c}$. The predictions in Ref.\cite{Wang:2017mqp} are two or
three times larger than ours since  in the two approaches the different pictures about the
inner structure are adopted.
We suggest the experimentalists to make more accurate measurements on the channels,
and the data would tell us which approach is closer to the reality.
Definitely, the theoretical studies on the double-heavy baryons are helpful
for getting a better understanding about the quark model and the non-perturbative QCD effects.
Especially, the scenarios adopted for investigating the double-charm baryons can be generalized
to study the baryons with $bb$ and $bc$ components which will be measured by the LHCb collaboration
and  other collaborations in the near future.
\section*{Acknowledgement}

This work is supported by the National Natural Science Foundation
of China (NNSFC) under the contract No. 11375128, 11675082,
11735010 and 11975165. We thank Prof. Wei Wang for helpful
discussions.

\appendix

\section{Semi-leptonic decays of  $\mathcal{B}_1\to
\mathcal{B}_2  l\bar\nu_l$ }

The helicity amplitudes are related to the form factors for
$\mathcal{B}_1\to \mathcal{B}_2 l\bar\nu_l$ through the following
expressions \cite{Korner:1991ph,Bialas:1992ny,Korner:1994nh}
 \beq
 H^V_{\frac{1}{2},0}&=&\frac{\sqrt{Q_-}}{\sqrt{q^2}}\left(
  \left(\Mb+\Mc\right)f_1-\frac{q^2}{\Mb}f_2\right),\non\\
 H^V_{\frac{1}{2},1}&=&\sqrt{2Q_-}\left(-f_1+
  \frac{\Mb+\Mc}{\Mb}f_2\right),\non\\
 H^A_{\frac{1}{2},0}&=&\frac{\sqrt{Q_+}}{\sqrt{q^2}}\left(
  \left(\Mb-\Mc\right)g_1+\frac{q^2}{\Mb}g_2\right),\non\\
 H^A_{\frac{1}{2},1}&=&\sqrt{2Q_+}\left(-g_1-
  \frac{\Mb-\Mc}{\Mb}g_2\right).
 \eeq
where $Q_{\pm}=2(P\cdot P'\pm \Mb\Mc)$ and $\Mb\, (\Mc)$
represents $M_{\Xi_{cc}}$ ($M_{\Xi_{c}}$). The amplitudes for the
negative helicities are obtained in terms of the relation
 \beq
 H^{V,A}_{-\lambda'-\lambda_W}=\pm H^{V,A}_{\lambda',\lambda_W},
  \eeq
where the upper (lower) index corresponds to V(A).
 The helicity
amplitudes are
 \beq
 H_{\lambda',\lambda_W}=H^V_{\lambda',\lambda_W}-
  H^A_{\lambda',\lambda_W}.
 \eeq
The helicities of the $W$-boson $\lambda_W$ can be either $0$ or
$1$, which correspond to the longitudinal and transverse
polarizations, respectively.  The longitudinally (L) and
transversely (T) polarized rates are
respectively\cite{Korner:1991ph,Bialas:1992ny,Korner:1994nh}
 \beq
 \frac{d\Gamma_L}{d\omega}&=&\frac{G_F^2|V_{cb}|^2}{(2\pi)^3}~
  \frac{q^2~p_c~\Mc}{12\Mb}\left[
  |H_{\frac{1}{2},0}|^2+|H_{-\frac{1}{2},0}|^2\right],\non\\
 \frac{d\Gamma_T}{d\omega}&=&\frac{G_F^2|V_{cb}|^2}{(2\pi)^3}~
  \frac{q^2~p_c~\Mc}{12\Mb}\left[
  |H_{\frac{1}{2},1}|^2+|H_{-\frac{1}{2},-1}|^2\right].
 \eeq
where $p_c$ is the momentum of $\mathcal{B}_2$ in the reset frame
of $\mathcal{B}_1$.

 The ratio of the longitudinal to
transverse decay rates $R$ is defined by
 \beq
 R=\frac{\Gamma_L}{\Gamma_T}=\frac{\int_1^{\omega_{\rm
     max}}d\omega~q^2~p_c\left[ |H_{\frac{1}{2},0}|^2+|H_{-\frac{1}{2},0}|^2
     \right]}{\int_1^{\omega_{\rm max}}d\omega~q^2~p_c
     \left[ |H_{\frac{1}{2},1}|^2+|H_{-\frac{1}{2},-1}|^2\right]}.
 \eeq

\section{$\mathcal{B}_1\to
\mathcal{B}_2 M$} In general, the transition amplitude of
$\mathcal{B}_1\to \mathcal{B}_2 M$ can be written as
 \beq
 {\cal M}(\mathcal{B}_1\to
\mathcal{B}_2 P)&=&\bar
  u_{\Lambda_c}(A+B\gamma_5)u_{\Lambda_b}, \non \\
 {\cal M}(\mathcal{B}_1\to
\mathcal{B}_2 V)&=&\bar
  u_{\Lambda_c}\epsilon^{*\mu}\left[A_1\gamma_{\mu}\gamma_5+
   A_2(p_c)_{\mu}\gamma_5+B_1\gamma_{\mu}+
   B_2(p_c)_{\mu}\right]u_{\Lambda_b},
 \eeq
where $\epsilon^{\mu}$ is the polarization vector of the final
vector or axial-vector mesons. Including the effective Wilson
coefficient $a_1=c_1+c_2/N_c$, the decay amplitudes in the
factorization approximation are \cite{Korner:1992wi,Cheng:1996cs}
 \beq
 A&=&\lambda f_P(\Mb-\Mc)f_1(M^2), \non \\
 B&=&\lambda f_P(\Mb+\Mc)g_1(M^2), \non\\
 A_1&=&-\lambda f_VM\left[g_1(M^2)+g_2(M^2)\frac{\Mb-\Mc}{\Mb}\right],
 \non\\
 A_2&=&-2\lambda f_VM\frac{g_2(M^2)}{\Mb},\non\\
 B_1&=&\lambda f_VM\left[f_1(M^2)-f_2(M^2)\frac{\Mb+\Mc}{\Mb}\right],
 \non\\
 B_2&=&2\lambda f_VM\frac{f_2(M^2)}{\Mb},
 \eeq
where $\lambda=\frac{G_F}{\sqrt 2}V_{cs}V_{q_1q_2}^*a_1$ and $M$
is the meson mass. Replacing  $P$, $V$ by $S$ and $A$ in the above
expressions, one can easily obtain similar expressions for scalar
and axial-vector mesons .

The decay rates of $\mathcal{B}_1\rightarrow \mathcal{B}_2P(S)$
and up-down asymmetries are\cite{Cheng:1996cs}
 \begin{eqnarray}
 \Gamma&=&\frac{p_c}{8\pi}\left[\frac{(\Mb+\Mc)^2-M^2}{\Mb^2}|A|^2+
  \frac{(\Mb-\Mc)^2-M^2}{\Mb^2}|B|^2\right], \non\\
 \alpha&=&-\frac{2\kappa{\rm Re}(A^*B)}{|A|^2+\kappa^2|B|^2},
 \end{eqnarray}
where $p_c$ is the $\mathcal{B}_2$ momentum in the rest frame of
$\mathcal{B}_1$. For $\mathcal{B}_1\rightarrow \mathcal{B}_2 V(A)$
decays, the decay rate and up-down asymmetries are
 \beq
 \Gamma&=&\frac{p_c (E_{\Lambda_c}+\Mc)}{4\pi\Mb}\left[
  2\left(|S|^2+|P_2|^2\right)+\frac{E^2}{M^2}\left(
  |S+D|^2+|P_1|^2\right)\right],\non\\
 \alpha&=&\frac{4M^2{\rm Re}(S^*P_2)+2E^2{\rm Re}(S+D)^*P_1}
  {2M^2\left(|S|^2+|P_2|^2 \right)+E^2\left(|S+D|^2+|P_1|^2
  \right) },
 \eeq
where $E$ is energy of the vector (axial vector) meson, and
 \begin{eqnarray}
  S&=&-A_1, \non\\
  P_1&=&-\frac{p_c}{E}\left(\frac{\Mb+\Mc}
  {E_{\Lambda_c}+\Mc}B_1+M_bB_2\right), \non \\
  P_2&=&\frac{p_c}{E_{\Lambda_c}+\Mc}B_1,\non\\
  D&=&-\frac{p^2_c}{E(E_{\Lambda_c}+\Mc)}(A_1- M_bA_2).
 \end{eqnarray}

\end{document}